\documentclass[onecolumn,10pt, cleanfoot]{asme2ej}

\usepackage{epsfig} 
\usepackage{arydshln}
\usepackage{caption}
\usepackage{helvet}

\DeclareCaptionLabelFormat{barlabel}{\textbf{#1~#2\space\textbar}}
\usepackage[hidelinks]{hyperref}
\usepackage{subfig}
\usepackage{amsmath}
\usepackage{float}

\usepackage{wrapfig}

\usepackage{bm}

\usepackage{braket}
\usepackage{amsmath,amssymb,amsfonts, float, braket, mathrsfs}    
\usepackage{graphicx, subfig, epsfig}                           
\usepackage{hyperref}        
\usepackage{amssymb}
\usepackage{graphicx}

\usepackage[backend=biber,style=nature]{biblatex}
\NewDocumentCommand{\figref}{m o}{
  \hyperref[#1]{Fig.~\ref*{#1}\IfValueT{#2}{#2}}}

\title{Electron Ptychography Reveals Correlated Lattice Vibrations at Atomic Resolution}

\author{
\parbox{\textwidth}{\centering Anton Gladyshev$^{\ast 1}$, Benedikt Haas$^1$, Thomas C. Pekin$^{1}$, Tara M. Boland$^{2,3}$, Marcel Schloz$^{1}$,Peter Rez$^4$ and  Christoph T. Koch$^1$} 
 \affiliation{$^1$ Department of Physics, Humboldt Universität zu Berlin \& Center for the Science of Materials Berlin, Berlin, Germany\\
  $^2$ School for Engineering of Matter Transport and Energy, Arizona State University, Tempe, AZ, United States\\
  $^3$ Computational Atomic-Scale Materials Design (CAMD), Technical University of Denmark, Kgs. Lyngby, Denmark\\
  $^4$ Department of Physics, Arizona State University, Tempe, AZ, United States\\
	$^\ast$  Email: gladysan@hu-berlin.de\\
            \today
	    }	
}

\begin{document}

\maketitle    
\begin{abstract}
{\it In this paper we introduce an electron ptychography reconstruction framework, CAVIAR – Correlated Atomic Vibration Imaging with sub-\AA ngstrom Resolution – that reveals an entirely new channel of information: spatial correlations in atomic displacements at the atomic scale. We show reconstructions of a symmetric $\Sigma$9 grain boundary in silicon from realistically simulated data and experimental data for hexagonal boron nitride. By reconstructing the object as an ensemble of multiple states we are able to observe correlations between movements of atoms in the range of 10-20 pm at room temperature in agreement with the expectation. Moreover, using only the masses of the atomic species and the temperature as input, we obtain average frequencies of  
$10.8\pm0.1$, $13.6\pm0.6$, $18.0\pm0.2$, $25.5\pm1.5$ THz for the longitudinal and transversal acoustic and optic phonons, respectively, in agreement with inelastic neutron scattering, albeit from just a few nm$^3$ volume. This ability to spatially resolve correlated atomic motion makes CAVIAR a unique tool to explore atom dynamics at the finest scale with the potential to be instrumental in the development of phononic devices, in studying phonon-based decoherence in quantum systems, or other emerging phonon-based applications.}
\end{abstract}

While capable of imaging the lateral positions of atomic columns constituting thin slabs of material, the achievable resolution of conventional electron imaging techniques in a transmission electron microscope (TEM) is very sensitive to instrumental imperfections: the partial coherence of the electron source, lens aberrations as well as mechanical and electronic instabilities of the microscope. Ptychography \cite{hoppe_1, hoppe_2, hoppe_3, hoppe_4, muller_deep_subangstrom} is a computational phase retrieval technique that, to some extent, can compensate for the imperfections of the equipment. Numerous quite different variations of this technique exist, e.g. Fourier and near-field ptychography \cite{fourier_ptycho, near_field_ptyhco}, and a variety of reconstruction schemes for them. The method used in this paper, "classical" far-field ptychography, recovers a complex transmission function of a specimen from a four-dimensional scanning transmission electron microscopy (4D-STEM) dataset containing transmitted intensities \cite{4d-stem} collected with a pixelated detector while illuminating overlapping areas of its surface with a convergent beam. 

\begin{figure*}[ht!]
\centering
\includegraphics[width=\textwidth]{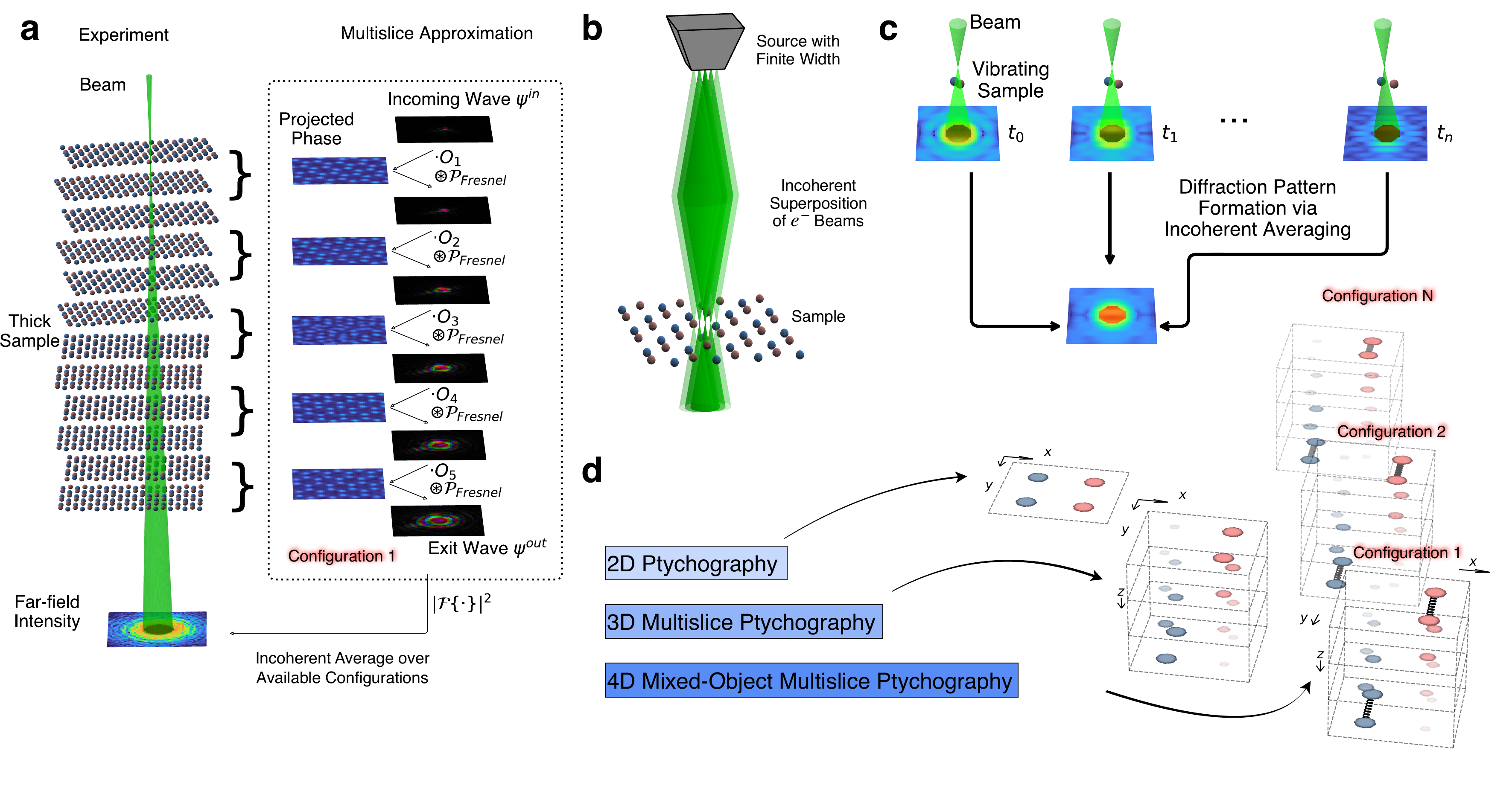}
    \caption{\textbf{Principles of numerical diffraction pattern formation in ptychography.} \textbf{a,} Schematic illustration of the  multislice 4D-STEM simulation for a thick sample. First, the beam propagation is split into multiple intervals (5 in the depicted case). The projected electrostatic potential from each of the intervals (slices) constitutes the phase of the corresponding transmission function $O_i$. The exit wavefront $\psi^{out}$ is obtained via sequential application of the transmission functions $O_i$ and Fresnel propagators $\mathcal{P}_{Fresnel}$ to the incoming wavefront $\psi^{in}$ \cite{Cowley1957_Multislice}. The diffraction pattern is calculated as the squared modulus of the Fourier transformed exit wave. In order to account for various types of incoherent scattering, the diffraction pattern is obtained as the sum over the scattered intensities from different configurations in the detector plane. Two types of variations in configuration are depicted in \textbf{b} and \textbf{c}. \textbf{b,} The effect of finite electron source size, creating a superposed ensemble of shifted beams where the occurrence probability of each beam is approximated to obey a Gaussian distribution. \textbf{c,} Principle of thermal diffuse scattering (TDS) caused by atomic movement. The diffraction patterns corresponding to quasi-static sample configurations \cite{loane_TDS_1991, muller_TDS_2001, Thibault_mixed_state} at multiple points in time $t_i$ are incoherently summed to obtain an expected intensity of a diffraction pattern with diffuse background. \textbf{d,} Electron ptychography reconstruction models with increasing complexity levels: 2D projected phase reconstruction, 3D multislice ptychography and our proposed reconstruction method – 4D mixed-object multislice ptychography, where the sequence of object states can be transformed to obtain correlations in variations of interatomic distances, schematically depicted as springs.}\label{fig: intro}
\end{figure*}
Ptychography has proven itself as a powerful tool for imaging of thin samples \cite{Nellist_1995, muller_deep_subangstrom}. Recently it was shown that adopting a multislice formalism makes it possible to resolve specimen features as fine as the blurring due to the vibrations of atoms \cite{muller_vibrations}. This publication has had greatly impacted the electron microscopy community and triggered a rapid rise in popularity of ptychography. But going beyond this limit was not possible, since a coherent specimen model as depicted in \figref{fig: intro}[a] cannot describe the formation of an incoherent diffraction pattern of the type  schematically shown in \figref{fig: intro}[c]. Previously there were attempts to include an incoherent sample model in experiments with laser \cite{maiden_mixed_obj} and X-ray illumination \cite{enders_mixed_obj, Thibault_mixed_state}, but applications to electron ptychography assumed only uncorrelated atomic vibrations \cite{Ziria_prb, Diederichs_2023}. Here we propose to utilize thermal diffuse scattering (TDS) \cite{muller_TDS_2001} as  a new source of information about correlated atomic movement rather than the ultimate limiting factor in achievable spatial resolution \cite{muller_vibrations}. This information can be retrieved using our proposed ptychographic reconstruction scheme, CAVIAR (Correlated Atomic Vibration Imaging with sub-\AA ngstrom Resolution), which is based on mixed-object electron ptychography. It combines a mixed-object formalism \cite{Thibault_mixed_state}, where the specimen is modeled as a statistical ensemble of states, with lattice Green’s function analysis \cite{Kong_phonon}, a technique well-established in molecular dynamics for quantifying vibrational correlations. The evolution from 2D ptychography \cite{ePIE} to 3D multislice ptychography \cite{chen_tcmulti, muller_vibrations} to CAVIAR is shown in \figref{fig: intro}[d]. With the retrieval of information about correlated movements of atoms (at atomic resolution), mixed-object ptychography becomes a complementary technique to vibrational electron energy-loss spectroscopy (EELS) in  scanning transmission electron microscopy (STEM)  \cite{gao_review, Haas_APL}. Vibrational STEM-EELS has been very successful in studying vibrational phenomena down to the atomic scale, such as vibrational states of single atoms \cite{Hage_EELS} or localized modes at grain boundaries \cite{haas_gb}. While vibrational EELS has inherent energy resolution, it can either be employed at atomic resolution without substantial momentum-resolution or momentum-resolved for mapping phonon dispersions but then with a spatial resolution insufficient for atomic resolution, as implied by Heisenberg's uncertainty principle. As mixed-object ptychography works fundamentally differently from EELS, it offers complementary information about the correlated motion of neighboring atoms at atomic resolution and should prove another useful tool for the study of atomic vibrations at the finest scale. As both techniques are based on STEM, they can, in principle, be performed using the same instrument, given the necessary hardware. 

The first step of CAVIAR is a gradient-based ptychographic reconstruction \cite{paper_rop, Wouter_2012, Wouter_2013, muller_deep_subangstrom, marcel_hbn, muller_vibrations, chen_tcmulti} which fits a transmission function of the specimen by determining the most likely \cite{rodenburg_noise}  parameters of a numerical model for a given set of experimentally measured intensities. Various aspects of the numerical model are depicted in \figref{fig: intro} and a detailed mathematical description can be found in the Supplementary Information (equations S1 -- S7). Upon convergence of the fit, one obtains a discrete complex transmission function of the specimen, which in our case has four dimensions: two lateral coordinates ($x$,$y$) spanning a plane perpendicular to the beam direction, an out-of-plane coordinate ($z$), parallel to the beam; and one dimension accommodating incoherent object modes ($n$). The phase of the transmission function $O(x,y,z,n)$ is directly  proportional to the projected electrostatic potential in slice $z$ \cite{muller_vibrations} of the $n$'th configuration of the object. The second step of CAVIAR is an extraction of the interatomic correlations from an unordered sequence of the retrieved object states $n$. To do so, we adopt a framework developed for the calculation of phonon dispersion curves from molecular dynamics simulations \cite{Kong_phonon}. The initial determination of atomic positions can be achieved either manually or, in more complex scenarios, using an object detection network (e.g. \cite{Yolo_main}). Subsequently, refining the atomic coordinates can be accomplished by calculating the center of mass within a small circular region around the initially determined atomic positions, using the retrieved phase values as weighting factors. In this way, for each unit cell $l$, basis atom $\kappa$, direction of displacement $\alpha$ ($x$,$y$,$z$) and reconstructed object state $n$ one can get real-space atomic positions $R_{l \kappa\alpha n}$ and displacements from the mean position \cite{Kong_phonon}:
\begin{equation}
    u_{l \kappa\alpha n}=R_{l \kappa\alpha n}- \frac{1}{N}\sum_{n=1}^N R_{l \kappa\alpha n}.
\end{equation}

Afterwards, the lattice Green’s function coefficient $G_{l \kappa\alpha l' \kappa' \alpha'}$ describing the correlations can be computed as a second moment of the displacements \cite{landau_lifshitz, Kong_phonon}:
\begin{equation}
    G_{l \kappa\alpha l' \kappa' \alpha'} = \frac{1}{N}\sum_{n=1}^{N}  u_{l \kappa\alpha n}  u_{l' \kappa'\alpha' n}.\label{eq: Green's_function}
\end{equation}

\begin{figure*}[ht!]
\centering
\includegraphics[width=0.9\textwidth]{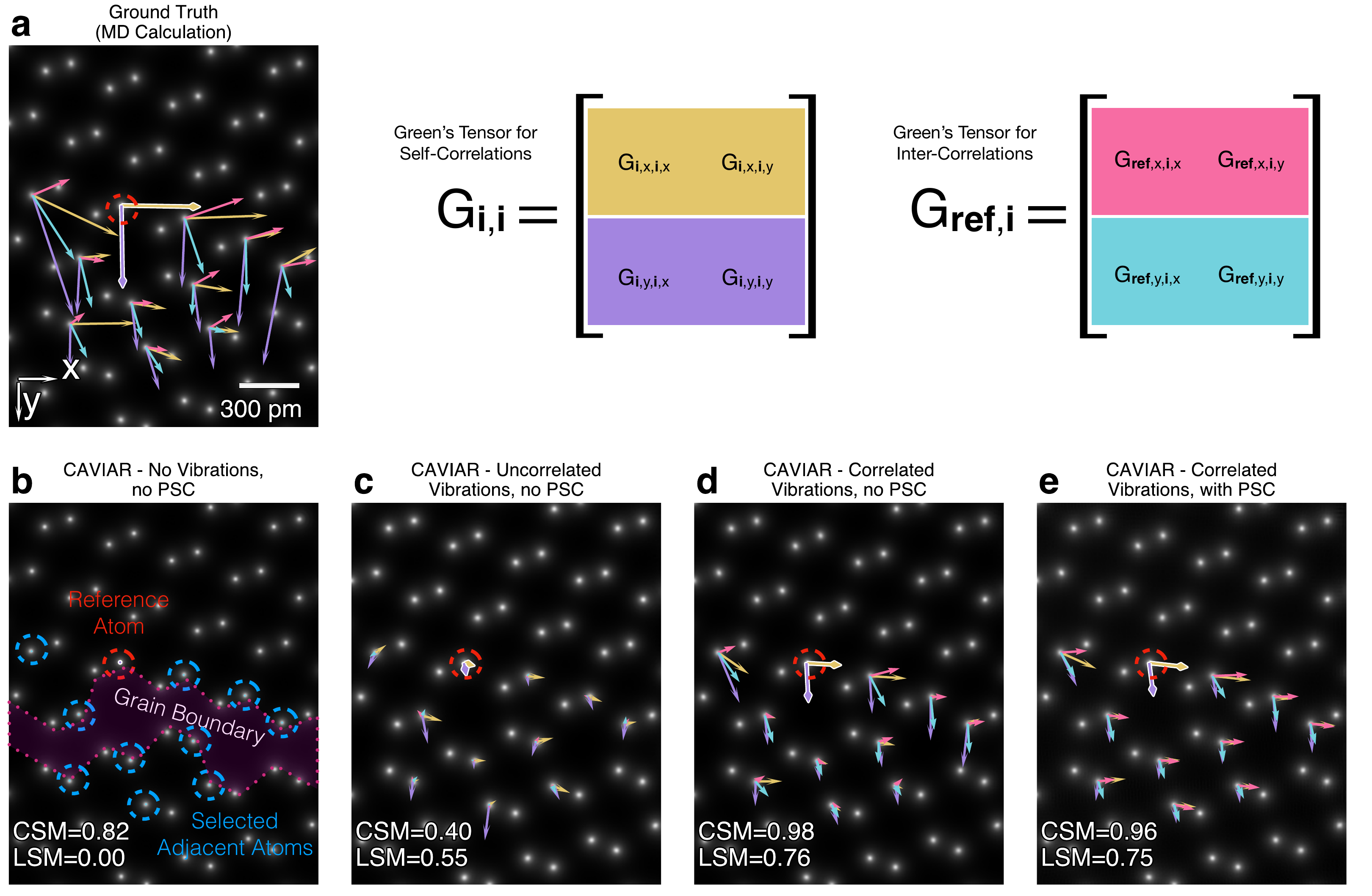}
    \caption{\textbf{Ptychographic reconstructions from realistically simulated data for various kinds of atomic motion and partial spatial coherence (PSC).} 
    \textbf{a,} Ground truth phase image from an MD simulation of a $\Sigma9$ grain boundary in silicon with arrows indicating various components of the Green's tensor defined in eq. \ref{eq: Green's_function}. A randomly picked reference atom is marked by a red circle in \textbf{a -- e}, its selected neighbors and the grain boundary are shown by a blue and magenta dashed lines in \textbf{b}, respectively. Colors of arrows show different correlation types: yellow/purple (self-correlations), pink/light blue (correlation with reference atom), two $2\times 2$ blocks of the full Green's tensor are shown schematically in the middle of the top row using the same colors. \textbf{b -- e,} Reconstructed phases and correlation matrices obtained from data simulated using different levels of approximation: 
    \textbf{b} Perfectly coherent dataset, no PSC — no correlations appear to have been reconstructed; 
    \textbf{c} Uncorrelated vibrations (Einstein model), no PSC — correlations between neighbors largely vanish; 
    \textbf{d} Correlated vibrations, no PSC — principal directions of correlations match the ground truth; 
    \textbf{e} Correlated vibrations, with PSC — the recovered correlations match the ground truth and the offset created by PSC is efficiently mitigated by multiple probe modes; 
     In the bottom of images \textbf{b -- e} we report values of length-similarity (LSM) and cosine-similarity (CSM) metrics (cf. eq. \ref{eq: lsm} and \ref{eq: csm} in the main text) between MD ground truth and retrieved correlation vectors. All reconstructions used 4 object slices, 20 object states. Reconstructions in \textbf{b -- d} were done with a single probe mode. In the reconstruction \textbf{e} 9 probe modes were used.}\label{fig: silicon_simulation}
\end{figure*}

To investigate the feasibility of CAVIAR, we started with simulated data to have well-controlled conditions and the ability to compare with the ground truth, before considering experimental data. As a test sample, we chose a symmetric $\Sigma$9 grain boundary in silicon, molecular dynamics (MD) simulations of phonon spectra of which \cite{rez_vibrations} agree perfectly well with experimental STEM-EELS data \cite{haas_gb}. Moreover, prior work has shown that atoms at the grain boundary exhibit distinct vibrational behavior compared to those in bulk regions due to differences in bonding environments \cite{haas_gb}. CAVIAR enables this phenomenon to be examined at the level of individual atoms. 

To generate a 4D-STEM dataset \cite{abtem} we used 30 time-snapshots (object configurations) of a 0.7 nm thick molecular dynamics (MD) super-cell containing four atomic layers. The accelerating voltage, convergence semi-angle, and scan-step were 200 kV,  30 mrad, and 50 pm, respectively. Averaging neighboring diffraction patterns with a Gaussian weighting was used to simulate the effect of partial spatial coherence with an effective source size (full width at half maximum of the Gaussian) of 47 pm. In total we generated four 4D-STEM datasets mimicking different degrees of incoherent scattering: 1) a perfectly coherent dataset based on a single MD snapshot, i.e. without thermal diffuse scattering and without additional partial spatial coherence; 2) a dataset based on the Einstein model of thermal vibration \cite{Einstein_1907, muller_TDS_2001, loane_TDS_1991}, i.e. uncorrelated displacements sampled from a normal distribution with standard deviation matched to the average one from the MD data and without additional partial spatial coherence; 3) a dataset based on MD data including correlated atomic movement without partial spatial coherence; 4) dataset  based on MD data with additional partial spatial coherence. Indicating the partial spatial coherence of the 4D-STEM data is important due to the fact that averaging over multiple slightly displaced probes is conceptually equivalent to averaging over multiple slightly displaced objects. Thus, when single probe mode is used in a reconstruction, partial spatial coherence, absorbed by the object states, introduces a positive offset to the $x$-$x$ and $y$-$y$ correlations of the atomic coordinates. The full Green’s tensors in Supplementary Fig. S2 reflect this effect, showing a uniform offset in the $x$-$x$ and $y$-$y$ correlations in the single probe mode reconstructions from the datasets corrupted by partial spatial coherence. Mitigation of this effect can be achieved in two ways: 1) the positive offset can be corrected in the recovered correlations, for example, in the same manner as background subtraction is performed in EELS spectra \cite{haas_gb}; 2) A ptychographic fit with single object mode and multiple probe modes \cite{muller_vibrations,  Thibault_mixed_state} can be performed prior to mixed-object and mixed-probe reconstruction in order to fit multiple probe states that will absorb the offset. In the reconstructions presented further we chose the second approach. All ptychographic fits from the Si grain boundary data were performed using 20 object states and  4 slices with a spacing of $\Delta z =0.19$ nm. In \figref{fig: silicon_simulation} we show both the reconstructed projected phase as well as the retrieved interatomic correlations. The reconstructions from data without additional partial spatial coherence were done with a single probe mode and the reconstruction in \figref{fig: silicon_simulation}[e] was done with 9 probe modes.

 Although there is no fundamental bound prohibiting a full 3D reconstruction, the $z$-resolution currently achievable with STEM-ptychography \cite{chen_tcmulti} is not sufficient to quantify the out-of-plane vibrations. Therefore, we limit our analysis  to the in-plane components of vibrations. For a system of $N$ atoms and two spatial coordinates, a retrieved Green's tensor depicting all correlations is a $2N \times 2N$ matrix. In order to visualize it in a convenient form, we first randomly picked a reference atom, shown in \figref{fig: silicon_simulation}[b] by the red circle, and some adjacent atoms circumscribed by the blue circles in the same image (for clarity of the figure only a few atoms were chosen). For a given pair of atoms we depict correlations as two vectors: the first one (coral) representing correlations of the reference atom's $x$ coordinate with the neighbor's $x$ and $y$ coordinates and the second one (cyan) representing correlations between the reference atom's $y$ coordinate with its neighbor's $x$ and $y$ coordinates. The yellow and purple arrows represent the atom's auto-correlation, which corresponds to the two contributions $\langle u_x^2 \rangle$ and $\langle u_y^2 \rangle$ of the equivalent isotropic displacement factor \cite{Fischer88_Uequiv}. 

\figref{fig: silicon_simulation}[a] shows the ground truth of the displacement correlations from the  MD snapshots. One can see that the correlations retrieved via CAVIAR (\figref{fig: silicon_simulation}[d, e]) are weaker than they should be, but the angles between the depicted arrows are nearly identical to the ground truth. Further, with our approach we can distinguish between various kinds of vibrations. In case of a coherent simulation (\figref{fig: silicon_simulation}[b]) we cannot detect any displacements at all and for the simulation with uncorrelated atom displacements (Einstein model, \figref{fig: silicon_simulation}[c]) we do not recover any correlations between neighbors. 

To quantitatively compare the correlations from the initial MD data and those extracted from the ptychographic reconstructions, we adopt two orthogonal metrics judging agreement in strength and direction: length similarity metric (LSM) and cosine-similarity metric (CSM) \cite{cos_metric_1}. For each atomic pair $i$, whose correlations are represented in \figref{fig: silicon_simulation} by a vector $V_i$ with two components (denoted $j$), we evaluate the similarity between the reconstructed ($Rec$) and ground truth ($GT$) correlations as follows:
\begin{align}
    LSM&=\sum_{ij} \sqrt{\frac{|V^{Rec}_{ij}|^2}{|V^{GT}_{ij}|^2}} \label{eq: lsm} \\
    CSM&=\frac{1}{N_{vectors}}\sum_{i=1}^{N_{vectors}}\frac{\sum_{j} V^{GT}_{ij}\cdot V^{Rec}_{ij}}{\sqrt{\sum_{j} \left(V^{GT}_{ij}\right)^2 }\cdot \sqrt{\sum_{j} \left(V^{Rec}_{ij}\right)^2 }} \label{eq: csm}.
\end{align}

The CSM metric is bounded between $-1$ and $1$, where $1$ means perfect similarity. The LSM metric is bounded from below by zero, meaning an absence of vibrations in a reconstruction.  LSM$=1$ also means a perfect similarity.  For the reconstruction from the dataset without partial spatial coherence (\figref{fig: silicon_simulation}[d]) the CSM value is the highest (0.98), showing an almost perfect match. When adding partial spatial coherence (\figref{fig: silicon_simulation}[e]), the mentioned positive offset in correlations reduces the CSM value only slightly to $0.96$, indicating that multiple probe modes efficiently eliminate the ambiguity. For the coherent dataset  and the one with uncorrelated TDS presented in \figref{fig: silicon_simulation}[b, c], the CSM values are substantially lower. The LSM metric allows to quantify the magnitude of retrieved vibrations, for a perfectly coherent dataset in \figref{fig: silicon_simulation}[b] the value is close to zero, indicating the absence atomic vibrations. Despite a good match in directions of correlations retrieved from correlated datasets, their magnitudes are approximately $25 \%$ lower than the ground truth. Overall, we conclude that we may underestimate the strength of correlations due to the resolution in these reconstructions being limited, as well as due to a limited number of object states, but we can explore the directionality and relative differences between neighboring bonds.

\begin{figure*}[ht!]
\centering
\includegraphics[width=\textwidth]{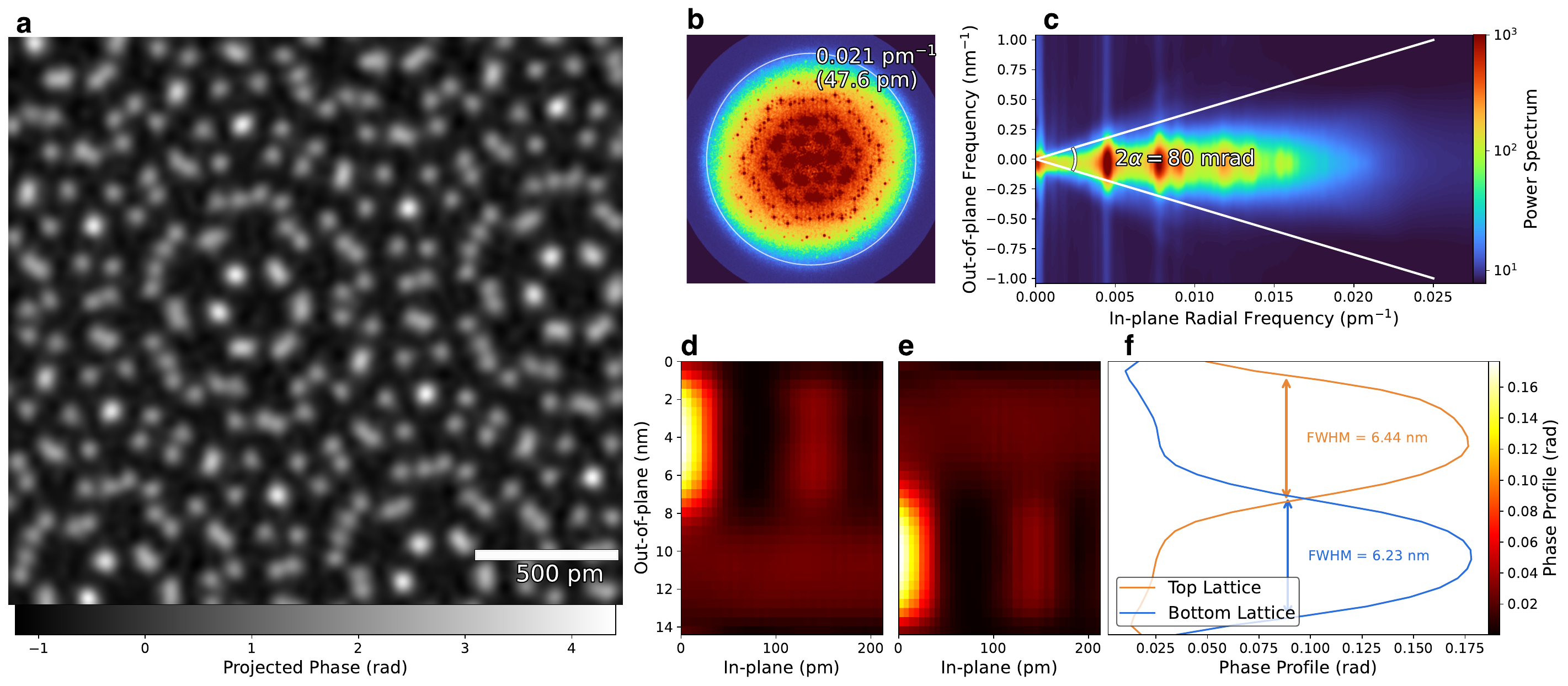}
    \caption{\textbf{Experimental ptychographic reconstruction from a 4D-STEM dataset of an approximately $15$ nm thick hBN bicrystal.} The specimens upper- and lower-halves (the beam propagates vertically) are twisted with respect to each other by approximately $11^\circ$. The reconstruction was carried out using 30 slices, 10 object states and 5 probe modes to account for partial spatial coherence and sample drift.  \textbf{a,} Reconstructed phase projected over all slices and averaged over all states. \textbf{b,} Sum of moduli squared of the Fast Fourier-transforms of all slices and object states indicating an information transfer up to 47.8 pm, way above the double-aperture radius limit (61 pm for an accelerating voltage of 60 kV and convergence semi-angle of 40 mrad). \textbf{c,} The azimuthal average of the 3D-Fourier moduli squared makes it possible to estimate the achieved resolution along the $z$-axis - approximately 2 nm for the first two rings of Bragg peaks. 
    \textbf{d, e,} Phase profiles averaged over all reconstructed atoms obtained by azimuthally  averaging phase profiles around the centers of atomic columns within the upper and lower slices of the reconstruction, respectively.  \textbf{f,} Phase profiles along $z$ through the centers of atomic columns clearly validating that two crystals have identical thicknesses within the precision limited by the depth resolution.}\label{fig: hbn_resoltuion}
\end{figure*}

After validating CAVIAR on simulated data we moved to an experimental 4D-STEM dataset of an approximately 15 nm thick hexagonal boron nitride (hBN) sample in [0001] direction, whose upper and lower halves were twisted by approximately $11^\circ$ relative to each other around said zone axis \cite{marcel_hbn}. The 4D-STEM dataset was acquired using a Nion HERMES microscope at an accelerating voltage of 60 kV, with convergence semi-angle of 40 mrad and a scan step of 30 pm. The reconstructed transmission function included  10 object states and 30 slices with a spacing of 0.48 nm. We further used 5 incoherent probe modes to account for partial spatial coherence of the beam and potentially occurring sample drift. 

In Supplementary Fig. S1, we demonstrate that even for a thin $\Sigma$9 grain boundary in silicon, multiple scattering dominates over incoherence arising from sample vibrations. Consequently, successful extraction of interatomic correlations via CAVIAR requires both sufficient slice sampling and adequate depth resolution. To validate that these conditions were met in our experiments, we begin by analyzing the achieved spatial resolution. \figref{fig: hbn_resoltuion}[a] shows the projected phase, summed over all 30 $z$-slices. In \figref{fig: hbn_resoltuion}[b] we show the  power spectrum incoherently averaged over object slices and states, which indicates an information transfer up to $d = 47.6$ pm. Dividing the achieved resolution by the illumination wavelength $\lambda=4.87$ pm gives us a value of $d = 9.78\ \lambda$. A ratio $d/\lambda < 10$  for a three dimensional potential reconstruction has so far been difficult to achieve \cite{muller_vibrations}. In order to estimate the depth resolution we azimuthally averaged a 3D Fourier transform of the phase over the in-plane spatial frequencies. The corresponding power spectrum is shown in \figref{fig: hbn_resoltuion}[c] and clearly indicates an information transfer within the full convergence angle of the beam (80 mrad) \cite{chen_tcmulti, muller_vibrations}. For the first two rings of hBN reflections it corresponds to approximately 2 nm of resolution along the beam-direction. In order to illustrate the separation between upper and lower lattices of the sample we computed average phase profiles as a function of distance $r$ to the center of the atomic columns in each of the reconstructed potential slices. In \figref{fig: hbn_resoltuion}[d -- e] these $r-z$ phase maps for the upper and lower lattices are shown. In \figref{fig: hbn_resoltuion}[f] the phase value at the center of the atomic columns is shown as a function of $z$ (line profiles along the left hand sides of \figref{fig: hbn_resoltuion}[d -- e]). The boundaries of each sub-lattice as well as an overlap region caused by the limited resolution along the $z$-axis can clearly be identified. 

\begin{figure*}[ht!]
\centering
\includegraphics[width=\textwidth]{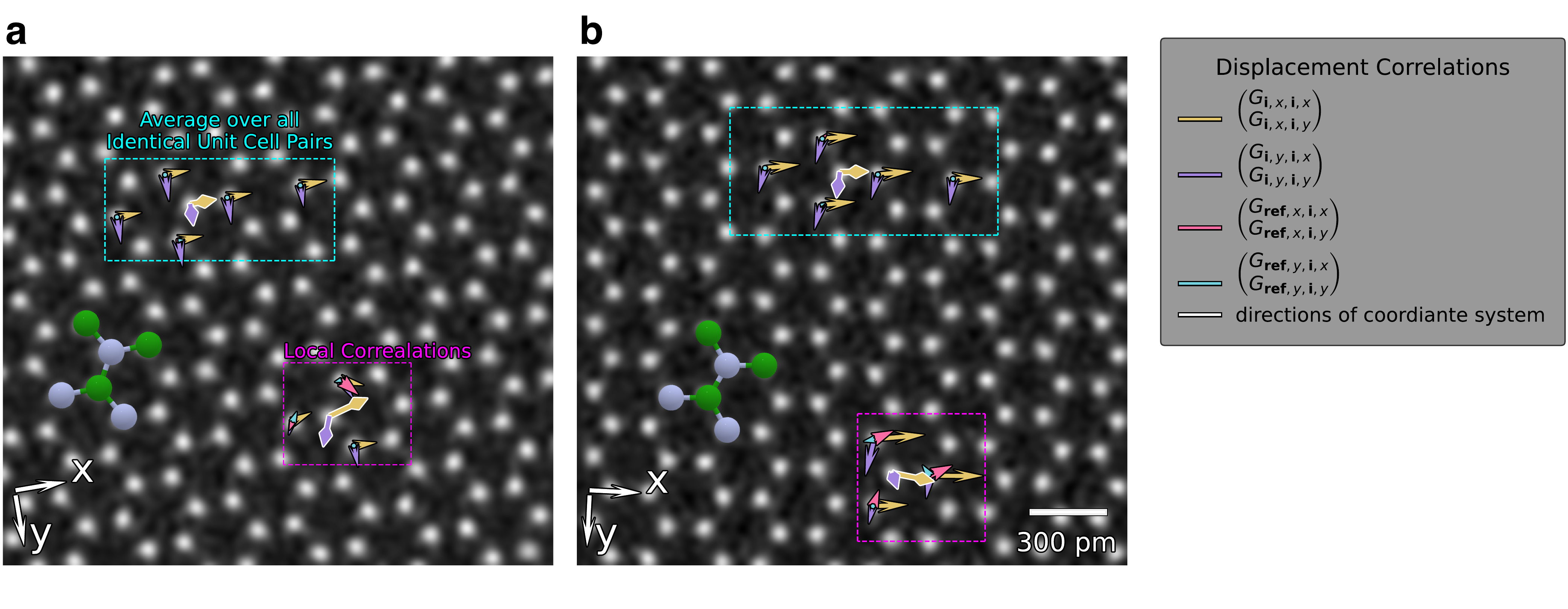}
    \caption{\textbf{Interatomic correlations recovered from experimental mixed-object ptychographic reconstruction of approximately $15$ nm thick bulk hBN crystal.}  \textbf{a, b,}  Projected phases of the reconstructed transmission function summed over all object states and 8 out of 30 slices selected from the $z$-ranges $z=1.9$ nm $\ldots 5.7$ nm and $z=8.6$ nm $\ldots 12.5$ nm, respectively. For each of the potential slices within those ranges the interatomic correlations were computed separately; their averages were computed afterwards. The depicted arrows represent various components of retrieved Green's tensors. Randomly picked reference atoms are shown by white outline, yellow and purple arrows show correlations of atoms with their own displacement along $x$- and $y$-axes. Pink and light blue show correlations of atoms with displacements of a reference atom along  $x$ or $y$ direction. Note that for the two stacked crystals the coordinate system was rotated to align the $x$-direction  with one of the three nearest neighbors. The $y$-axis was then defined to be orthogonal to it. 
    Directions of each coordinate system are shown  with white arrows in the lower left corners of each panel as well as the unit cells.
    It is important to mention that individual atoms may vibrate quite differently, but on average they are expected to behave similarly. In the magenta colored dashed boxes we show correlations corresponding to a particular atomic column (and its neighbors) and in turquoise boxes we show correlations averaged over all identical unit-cell pairs (i.e. with the same differences $l-l'$ and  $\kappa-\kappa'$). Note that principal directions of averaged correlations are identical in \textbf{a} and \textbf{b}.
    }\label{fig: hbn_corrs}
\end{figure*}

The depth resolution analysis above confirms that multiple scattering effects are accounted for in the reconstruction, enabling further analysis of interatomic correlations. For each of the 30 slices and 10 states we extracted atom positions and corresponding displacements. In Figure \figref{fig: hbn_corrs}[a, b] projected phases and interatomic correlations averaged in $z$-regions from 1.9 nm up to 5.7 nm and from 8.6 nm up to 12.5 nm, respectively, are shown. For each of the two lattices we evaluated correlations in coordinate systems aligned with the respective local lattice frame. In the magenta dashed boxes we show correlations corresponding to a particular atom column (and its neighbors) and in the turquoise boxes we show correlations averaged over all identical unit-cell pairs. The correlation vectors, overlaid at atomic sites, reveal similar spatial structures and orientations in both layers, despite their relative rotation. This confirms that the recovered correlations are physically meaningful and lattice-intrinsic. We further compute a Debye-Waller B-factor by summing self correlations ($x$-$x$ and $y$-$y$) and multiplying the sum by $8\pi^2$. For the upper and lower lattices we get $3\cdot 10^3$ pm$^2$ and $3.5\cdot 10^3$ pm$^2$ respectively, which are approximately 6 times lower than the reported value of $2\cdot 10^4$ pm$^2$ obtained from powder X-ray diffraction \cite{hbn_uiso}. 
Such a discrepancy cannot simply be explained by a lack of $z$-vibrations in our reconstruction, as the $z$-resolution of 2 nm in this ptychographic reconstruction is clearly not sufficient to detect fluctuations along this axis. 
Although for a 2D material, such as hBN, where the layers are bound only by the Van der Waals forces to one another, the $z$-vibrations might significantly increase the isotropic squared displacements, two additional limiting factors are responsible for reducing the amplitude of reconstructed correlations and the remaining broadening of the images of projected atom positions within the layers:
1) the $z$-distance between atomic layers in hBN is 0.33 nm, which means that per 0.48 nm thin potential slice there are on average 1.4 layers of atoms, the positions of which are averaged over, and 2) the limited number of only 10 object states may not be sufficient to represent all of the TDS and Debye-Waller factor by an incoherent sum of diffraction patterns from statically displaced atom configurations. 

Still, despite an underestimated correlation strength, their directions provide  direct access to the dynamical matrix of the crystal. Following the procedure described in \cite{Kong_phonon}, we compute Fourier-transformed displacements by summing over all $L$ available unit cells $l$:
\begin{equation}
    \tilde{u}_{\kappa\alpha}(q)=\frac{1}{\sqrt{L}}\sum_{l=1}^L u_{l\kappa\alpha} \times e^{-iq r_l}.
\end{equation}
By assuming boron's and nitrogen's harmonically averaged atomic mass of $12.2$ a.m.u. and a temperature of $300$ K we obtain the dynamical matrix \cite{Kong_phonon, landau_lifshitz} via:
\begin{equation}
    D_{\kappa\alpha \kappa'\beta} (q)=\frac{k_B T}{m} \left[ \langle\tilde{u}_{\kappa\alpha}(q)\tilde{u}^*_{\kappa'\beta}(q)\rangle \right]^{-1}_{\kappa\alpha \kappa'\beta}, 
\end{equation}
where $\kappa$, $\kappa'$ indicate one of two basis atoms and $\alpha$, $\beta$ indicate the displacement directions ($x$ and $y$). Computing the square root of eigenvalues of this matrix directly results in the phonon-dispersion curves. Although the 59 unit cells available in our scanned field of view are not sufficient to examine the curves in detail, we compute an average frequency along the $\Gamma-K-M-\Gamma$ path. We obtain $10.7$ THz, $14.1$ THz,  $17.7$ THz and $24.0$ THz for transverse acoustic, longitudinal acoustic, transverse optical, longitudinal optical branches in the upper lattice. For the bottom lattice we get  $10.8$ THz, $13.0$ THz, $18.2$ THz, and $27.0$ THz for the same four branches. The reported theoretical values are approximately 1.5 times higher: $14.3$ THz, $22.9$ THz, $39.5$ THz and $39.6$ THz \cite{hbn-dispersion}. The absolute values of the mode frequencies are heavily affected by the underestimated correlation strength, but they allow to place frequencies of two acoustic and two optical branches in the correct range for a complementary Vibrational STEM-EELS measurements \cite{Yan_EELS, Hage_EELS}. Further, unlike EELS, mixed-object ptychography yields directions of correlated vibrations at the level of a single atom.

Before summarizing above results we would like to point out that the experimental discovery of thermal streaks in electron diffraction patterns \cite{honjo1962proc, honjo1964diffuse} is as old as the idea of ptychography \cite{hoppe_1, hoppe_2, hoppe_3, hoppe_4}. The impact of lattice vibrations on electron diffraction was heavily investigated both experimentally and theoretically during the last century (e.g. \cite{jm_zuo_1990}) and is still a hot research topic today (e.g. \cite{PhysRevB.100.144305}). It has already been pointed out that the faint structure in the diffuse Kikuchi diffraction intensity between the Bragg spots of electron diffraction patterns makes it possible to distinguish between correlated (consistent with the details of the phonon dispersion curve) and uncorrelated (according to the Einstein model) lattice vibrations \cite{muller_TDS_2001}.  
Utilizing the computational power and recent advancements in ptychographic reconstruction algorithms, we have extended electron ptychography beyond static imaging by introducing the CAVIAR reconstruction framework capable of resolving spatial correlations in atomic displacements. By coupling a statistical object model \cite{Thibault_mixed_state} with lattice Green’s function analysis \cite{Kong_phonon}, we recover correlated atomic vibrations directly from experimental 4D-STEM data initially having no energy resolution. We report on deep sub-\AA ngstrom spatial resolution and its ratio to the illumination wavelength of $d/\lambda = 9.78$. Simulations presented above reveal that CAVIAR distinguishes between different kinds of lattice vibrations including correlated, uncorrelated, and absent motion, while experimental validation on a twisted hBN bicrystal demonstrates the ability to extract real-space correlation patterns and infer local phonon dispersion. This approach opens a new pathway for imaging lattice dynamics and thermal disorder, offering a complementary technique to vibrational STEM EELS \cite{Hage_EELS, Yan_EELS}.

\section*{Acknowledgement}
A.G., B.H. and C.T.K. acknowledge financial support by the Deutsche Forschungsgemeinschaft (DFG) in project no. 182087777 (CRC951) and project no. 414984028 (CRC1404). Authors thank Niklas Dellby and Paul Zeiger for helpful discussions and Dana O. Byrne and Jim Ciston for providing the hBN sample. A.G. thanks Meng Zhao for providing an encouraging environment throughout the development of the ptychographic code.

\section*{Author contributions}
A.G. performed all ptychographic reconstructions and data analysis. T.C.P., C.T.K., and B.H. designed the experiment; T.C.P. acquired the data, and B.H. carried out non-rigid registration. A.G., C.T.K., and P.R. developed the concepts of data analysis. M.S. contributed by correcting distortions introduced by the energy filter. T.B. and P.R. performed the molecular dynamics simulations of the silicon grain boundary. All authors discussed the results and implications throughout the investigation and approved the final version of the manuscript.

\printbibliography

@article{muller_vibrations,
  author    = {Zhen Chen and Yi Jiang and Yu-Tsun Shao and Megan E. Holtz and Michal Odstrčil and Manuel Guizar-Sicairos and Isabelle Hanke and Steffen Ganschow and Darrell G. Schlom and David A. Muller},
  title     = {Electron ptychography achieves atomic-resolution limits set by lattice vibrations},
  journal   = {Science},
  volume    = {372},
  number    = {6544},
  pages     = {826--831},
  year      = {2021},
  month     = {05},
  doi       = {10.1126/science.abg2533}
}

@article{chen_tcmulti,
	Author = {Dong, Zehao and Zhang, Yang and Chiu, Chun-Chien and Lu, Sicheng and Zhang, Jianbing and Liu, Yu-Chen and Liu, Suya and Yang, Jan-Chi and Yu, Pu and Wang, Yayu and Chen, Zhen},
	Da = {2025/01/31},
	Doi = {10.1038/s41467-025-56499-1},
	Journal = {Nature Communications},
	Number = {1},
	Pages = {1219},
	Title = {Sub-nanometer depth resolution and single dopant visualization achieved by tilt-coupled multislice electron ptychography},
	Ty = {JOUR},
	Volume = {16},
	Year = {2025},
	}

@article{hbn_uiso,
author = "Kurakevych, Oleksandr O. and Solozhenko, Vladimir L.",
title = "{Rhombohedral boron subnitride, B${\sb 13}$N${\sb 2}$, by X-ray powder diffraction}",
journal = "Acta Crystallographica Section C",
year = "2007",
volume = "63",
number = "9",
pages = "i80--i82",
month = "09",
doi = {10.1107/S0108270107037353},
url = {https://www.ccdc.cam.ac.uk/structures/Search?Ccdcid=1500064}
}

@unpublished{hbn-dispersion,
      title={Thermal expansion in 2D honeycomb structures: Role of transverse phonon modes}, 
      author={Sarita Mann and V. K. Jindal},
      year={2016},
Month = {06},
      eprint={1606.07656},
      archivePrefix={arXiv},
      primaryClass={cond-mat.mes-hall},
    Doi = {10.48550/arXiv.1606.07656},
 
}

@article{Haas_APL,
    author = {Haas, Benedikt and Koch, Christoph T. and Rez, Peter},
    title = {Perspective on atomic-resolution vibrational electron energy-loss spectroscopy},
    journal = {Applied Physics Letters},
    volume = {125},
    number = {15},
    pages = {150502},
    year = {2024},
    month = {10},   
    doi = {10.1063/5.0231688},
}

@article{l-bfgs,
author = "Liu, {Dong C.} and Jorge Nocedal",
year = "1989",
month = aug,
doi = "10.1007/BF01589116",
language = "English (US)",
volume = "45",
pages = "503--528",
journal = "Mathematical Programming",
issn = "0025-5610",
publisher = "Springer-Verlag GmbH and Co. KG",
number = "1-3",
}

@article{haas_gb,
  author    = {Benedikt Haas and Tara M. Boland and Christian Elsässer and Arunima K. Singh and Katia March and Juri Barthel and Christoph Koch and Peter Rez},
  title     = {Atomic-Resolution Mapping of Localized Phonon Modes at Grain Boundaries},
  journal   = {Nano Letters},
  volume    = {23},
  number    = {13},
  pages     = {5379--5386},
  year      = {2023},
  month     = {06},
  doi       = {10.1021/acs.nanolett.3c01089}
}

@article{4d-stem,
  author    = {Colin Ophus},
  title     = {Four-Dimensional Scanning Transmission Electron Microscopy (4D-STEM): From Scanning Nanodiffraction to Ptychography and Beyond},
  journal   = {Microscopy and Microanalysis},
  volume    = {25},
  number    = {3},
  pages     = {563--582},
  year      = {2019},
  month     = {06},
  doi       = {10.1017/S1431927619000497}
}

@article{hoppe_1,
  author    = {Walter Hoppe},
  title     = {Beugung im inhomogenen primärstrahlwellenfeld. I. Prinzip einer phasenmessung von elektronenbeugungsinterferenzen},
  journal   = {Acta Crystallographica Section A: Crystal Physics, Diffraction, Theoretical and General Crystallography},
  volume    = {25},
  number    = {4},
  pages     = {495--501},
  year      = {1969},
  month     = {07},
  doi       = {10.1107/S0567739469001045}
}

@article{hoppe_2,
  author    = {Walter Hoppe and G. Strube},
  title     = {Beugung in inhomogenen primärstrahlenwellenfeld. II. lichtoptische analogieversuche zur phasenmessung von gitterinterferenzen},
  journal   = {Acta Crystallographica Section A: Crystal Physics, Diffraction, Theoretical and General Crystallography},
  volume    = {25},
  number    = {4},
  pages     = {501--508},
  year      = {1969},
  month     = {07},
  doi       = {10.1107/S0567739469001057}
}

@article{hoppe_3,
  author    = {Walter Hoppe},
  title     = {Beugung im inhomogenen Primärstrahlwellenfeld. III. Amplituden-und Phasenbestimmung bei unperiodischen Objekten},
  journal   = {Acta Crystallographica Section A: Crystal Physics, Diffraction, Theoretical and General Crystallography},
  volume    = {25},
  number    = {4},
  pages     = {508--514},
  year      = {1969},
  month     = {07},
  doi       = {10.1107/S0567739469001069}
}

@article{hoppe_4,
  author    = {Reiner Hegerl and Walter Hoppe},
  title     = {Dynamische theorie der kristallstrukturanalyse durch elektronenbeugung im inhomogenen primärstrahlwellenfeld},
  journal   = {Berichte der Bunsengesellschaft für physikalische Chemie},
  volume    = {74},
  number    = {11},
  pages     = {1148--1154},
  year      = {1970},
  month     = {11},
  doi       = {10.1002/bbpc.19700741120}
}

@article{maiden_mixed_obj,
  author    = {Peng Li and Tega Edo and Darren Batey and John Rodenburg and Andrew Maiden},
  title     = {Breaking ambiguities in mixed state ptychography},
  journal   = {Optics express},
  volume    = {24},
  number    = {9},
  pages     = {9038--9048},
  year      = {2016},
  month     = {04},
  doi       = {10.1364/OE.24.009038}
}

@article{Thibault_mixed_state,
  author    = {Pierre Thibault and Andreas Menzel},
  title     = {Reconstructing state mixtures from diffraction measurements},
  journal   = {Nature},
  volume    = {494},
  number    = {7435},
  pages     = {68--71},
  year      = {2013},
  month     = {02},
  doi       = {10.1038/nature11806}
}

@article{lsq_ml,
  author    = {Michal Kronenberg and Andreas Menzel and Manuel Guizar-Sicairos},
  title     = {Iterative least-squares solver for generalized maximum-likelihood ptychography},
  journal   = {Optics express},
  volume    = {26},
  number    = {3},
  pages     = {3108--3123},
  year      = {2018},
  month     = {02},
  doi       = {10.1364/OE.26.003108}
}

@article{maiden_positions,
  author    = {Andrew Maiden and Martin Humphry and Michael Sarahan and B. Kraus and John Rodenburg},
  title     = {An annealing algorithm to correct positioning errors in ptychography},
  journal   = {Ultramicroscopy},
  volume    = {120},
  pages     = {64--72},
  year      = {2012},
  month     = {06},
  doi       = {10.1016/j.ultramic.2012.06.001}
}

@article{ePIE,
  author    = {Maiden, A.M. and Rodenburg, J.M.},
  title     = {An improved ptychographical phase retrieval algorithm for diffractive imaging},
  journal   = {Ultramicroscopy},
  volume    = {109},
  number    = {10},
  pages     = {1256--1262},
  year      = {2009},
  month     = {10},
  doi       = {10.1016/j.ultramic.2009.05.012}
}

@article{paper_rop,
  author = {Schloz, Marcel and Pekin, Thomas and Chen, Zhen and Broek, Wouter and Muller, David and Koch, Christoph},
  title = {Overcoming information reduced data and experimentally uncertain parameters in ptychography with regularized optimization},
  journal={Optics express},
  volume={28},
  number={19},
  pages = {28306--28323},
  year = {2020},
  month     = {09},
  doi       = {10.1364/OE.28.028306},
}

@article{ADORYM,
  title={Adorym: A multi-platform generic X-ray image reconstruction framework based on automatic differentiation},
  author={Du, Ming and Kandel, Saugat and Deng, Junjing and Huang, Xiaojing and Demortiere, Arnaud and Nguyen, Tuan Tu and Tucoulou, Remi and De Andrade, Vincent and Jin, Qiaoling and Jacobsen, Chris},
  journal={Optics express},
  volume={29},
  number={7},
  pages={10000--10035},
  year={2021},
  month     = {03},
  doi  = {10.1364/OE.429859}
}

@article{tilt_correction,
  author  = {Sha, Haozhi and Cui, Jizhe and Yu, Rong},
  title   = {Deep sub-angstrom resolution imaging by electron ptychography with misorientation correction},
  journal = {Science Advances},
  volume  = {8},
  number  = {22},
  pages   = {eabn2275},
  year    = {2022},
  month   = {05},
  doi     = {10.1126/sciadv.abn2275}
}

@article{Ziria_prb,
  title = {Thermal vibrations in the inversion of dynamical electron scattering},
  author = {Herdegen, Ziria and Diederichs, Benedikt and M\"uller-Caspary, Knut},
  journal = {Phys. Rev. B},
  volume = {110},
  issue = {6},
  pages = {064102},
  numpages = {17},
  year = {2024},
  month = {08},
  doi = {10.1103/PhysRevB.110.064102},
}

@article{Wouter_2012,
  title = {Method for Retrieval of the Three-Dimensional Object Potential by Inversion of Dynamical Electron Scattering},
  author = {Van den Broek, Wouter and Koch, Christoph T.},
  journal = {Phys. Rev. Lett.},
  volume = {109},
  issue = {24},
  pages = {245502},
  numpages = {5},
  year = {2012},
  month = {11},
  doi = {10.1103/PhysRevLett.109.245502},
}

@article{Wouter_2013,
  title = {General framework for quantitative three-dimensional reconstruction from arbitrary detection geometries in TEM},
  author = {Van den Broek, Wouter and Koch, Christoph T.},
  journal = {Phys. Rev. B},
  volume = {87},
  issue = {18},
  pages = {184108},
  numpages = {11},
  year = {2013},
  month = {05},
  doi = {10.1103/PhysRevB.87.184108},
}

@InProceedings{Yolo_main,
author = {Redmon, Joseph and Divvala, Santosh and Girshick, Ross and Farhadi, Ali},
title = {You Only Look Once: Unified, Real-Time Object Detection},
booktitle = {Proceedings of the IEEE Conference on Computer Vision and Pattern Recognition (CVPR)},
month = {06},
year = {2016}
}

@article{nonrigid,
  title={Increasing spatial fidelity and SNR of 4D-STEM using multi-frame data fusion},
  author={O'Leary, Colum M and Haas, Benedikt and Koch, Christoph T and Nellist, Peter D and Jones, Lewys},
  journal={Microscopy and Microanalysis},
  volume={28},
  number={4},
  pages={1417--1427},
  year={2022},
  publisher={Oxford University Press}
}

@book{landau_lifshitz,
  author    = {L. D. Landau and E. M. Lifshitz},
  title     = {Statistical Physics, Part 1},
  series    = {Course of Theoretical Physics},
  volume    = {5},
  edition   = {3},
  publisher = {Pergamon Press},
  year      = {1980},
  address   = {Oxford},
}

@article{Diederichs_2023,
  author    = {Diederichs, Benedikt and Herdegen, Ziria and Strauch, Achim and Filbir, Frank and M{\"u}ller-Caspary, Knut},
  title     = {Exact inversion of partially coherent dynamical electron scattering for picometric structure retrieval},
  journal   = {Nature Communications},
  volume    = {15},
  number    = {1},
  pages     = {101},
  year      = {2024},
  month     = {01},
  doi       = {10.1038/s41467-023-44268-x},
}

@article{Kong_phonon,
  author = {Kong, Ling},
  year = {2011},
  month = {10},
  pages = {2201-2207},
  title = {Phonon dispersion measured directly from molecular dynamics simulations},
  volume = {182},
  journal = {Computer Physics Communications},
  doi = {10.1016/j.cpc.2011.04.019}
}

@article{Wirtinger,
  author    = {Wilhelm Wirtinger},
  title     = {Zur formalen Theorie der Funktionen von mehr komplexen Veränderlichen},
  journal   = {Mathematische Annalen},
  volume    = {97},
  pages     = {357--375},
  year      = {1927},
  month     = {11},
  doi       = {10.1007/BF01447872}
}

@article{muller_TDS_2001,
  author    = {David A. Muller and Byard Edwards and Earl J. Kirkland and John Silcox},
  title     = {Simulation of thermal diffuse scattering including a detailed phonon dispersion curve},
  journal   = {Ultramicroscopy},
  volume    = {86},
  number    = {3-4},
  pages     = {371--380},
  year      = {2001},
  month     = {02},
  doi       = {10.1016/S0304-3991(00)00128-5}
}

@article{loane_TDS_1991,
  author    = {R. F. Loane and P. Xu and J. Silcox},
  title     = {Thermal Vibrations in Convergent-Beam Electron Diffraction},
  journal   = {Acta Crystallographica Section A},
  volume    = {47},
  number    = {3},
  pages     = {267--278},
  year      = {1991},
  month     = {05},
  doi       = {10.1107/S0108767391000375}
}

@article{Einstein_1907,
	Author = {Einstein, A. },
	Booktitle = {Annalen der Physik},
	Doi = {https://doi.org/10.1002/andp.19063270110},
	Journal = {Annalen der Physik},
	Month = {01},
    Year= {1907},
	Number = {1},
	Pages = {180--190},
	Publisher = {John Wiley \& Sons, Ltd},
	Title = {Die Plancksche Theorie der Strahlung und die Theorie der spezifischen W{\"a}rme},
	Ty = {JOUR},
	Volume = {327},
	Year = {1907},
}

@article{abtem,
  author    = {Jacob Madsen and Toma Susi},
  title     = {The abTEM code: transmission electron microscopy from first principles},
  journal   = {Open Research Europe},
  volume    = {1},
  pages     = {24},
  year      = {2021},
  month     = {03},
  doi       = {10.12688/openreseurope.13015.1}
}

@article{muller_deep_subangstrom,
  author    = {Yi Jiang and Zhen Chen and Yimo Han and Pratiti Deb and Hui Gao and Saien Xie and Prafull Purohit and Mark W. Tate and Jiwoong Park and Sol M. Gruner and Veit Elser and David A. Muller},
  title     = {Electron ptychography of 2D materials to deep sub-{\aa}ngstr{\"o}m resolution},
  journal   = {Nature},
  volume    = {559},
  pages     = {343--349},
  year      = {2018},
  month     = {07},
  doi       = {10.1038/s41586-018-0298-5}
}

@article{fourier_ptycho,
  author    = {Guoan Zheng and Cheng Shen and Shaowei Jiang and Pengming Song and Changhuei Yang},
  title     = {Concept, implementations and applications of Fourier ptychography},
  journal   = {Nature Reviews Physics},
  volume    = {3},
  pages     = {207--223},
  year      = {2021},
  month     = {02},
  doi       = {10.1038/s42254-021-00280-y}
}

@article{near_field_ptyhco,
    author = {You, Shengbo and Lu, Peng-Han and Schachinger, Thomas and Kovács, András and Dunin-Borkowski, Rafal E. and Maiden, Andrew M.},
    title = {Lorentz near-field electron ptychography},
    journal = {Applied Physics Letters},
    volume = {123},
    number = {19},
    pages = {192406},
    year = {2023},
    month = {11},
    doi = {10.1063/5.0169788}
}

@article{rez_vibrations,
  author    = {Peter Rez and Tara M. Boland and Christian Els{\"a}sser and Arunima K. Singh},
  title     = {Localized Phonon Densities of States at Grain Boundaries in Silicon},
  journal   = {Microscopy and Microanalysis},
  volume    = {28},
  pages     = {1--8},
  year      = {2022},
  month     = {03},
  doi       = {10.1017/S143192762200040X}
}

@article{marcel_3D_recon,
  author    = {Hamish G. Brown and Philipp M. Pelz and Shang-Lin Hsu and Zimeng Zhang and Ramamoorthy Ramesh and Katherine Inzani and Evan Sheridan and Sin{\'e}ad M. Griffin and Marcel Schloz and Thomas C. Pekin and Christoph T. Koch and Scott D. Findlay and Leslie J. Allen and Mary C. Scott and Colin Ophus and Jim Ciston},
  title     = {A Three-Dimensional Reconstruction Algorithm for Scanning Transmission Electron Microscopy Data from a Single Sample Orientation},
  journal   = {Microscopy and Microanalysis},
  volume    = {28},
  pages     = {1--9},
  year      = {2022},
  month     = {06},
  doi       = {10.1017/S1431927622012090}
}

@article{marcel_hbn,
    author = {Schloz, Marcel and Pekin, Thomas C and Brown, Hamish G and Byrne, Dana O and Esser, Bryan D and Terzoudis-Lumsden, Emmanuel and Taniguchi, Takashi and Watanabe, Kenji and Findlay, Scott D and Haas, Benedikt and Ciston, Jim and Koch, Christoph T},
    title = {Improved Three-Dimensional Reconstructions in Electron Ptychography through Defocus Series Measurements},
    journal = {Microscopy and Microanalysis},
    volume = {31},
    number = {1},
    pages = {ozae110},
    year = {2024},
    month = {11},
    doi = {10.1093/mam/ozae110}
    }

@inproceedings{cupy_learningsys2017,
  author    = {Okuta, Ryosuke and Unno, Yuya and Nishino, Daisuke and Hido, Shohei and Loomis, Crissman},
  title     = {CuPy: A NumPy-Compatible Library for NVIDIA GPU Calculations},
  booktitle = {Proceedings of the Workshop on Machine Learning Systems (LearningSys) at NeurIPS},
  year      = {2017},
  url       = {https://cupy.dev}
}

@phdthesis{enders_mixed_obj,
  author = {Enders, Björn},
  title  = {Development and Application of Decoherence Models in Ptychographic Diffraction Imaging},
  school = {Technische Universität München},
  year   = {2016}
}

@article{honjo1962proc,
  author  = {Honjo, G.},
  title   = {Proc. Int. Conf. Mag. and Cryst. 1961, Kyoto, II},
  journal = {Journal of the Physical Society of Japan},
  volume  = {17},
  number  = {B-II},
  pages   = {277},
  year    = {1962}
}

@article{honjo1964diffuse,
  author  = {Honjo, Goro and Kodera, Shiro and Kitamura, Norihisa},
  title   = {Diffuse streak diffraction patterns from single crystals I. General discussion and aspects of electron diffraction diffuse streak patterns},
  journal = {Journal of the Physical Society of Japan},
  volume  = {19},
  number  = {3},
  pages   = {351--367},
  year    = {1964},
  month   = {03},
  doi     = {10.1143/JPSJ.19.351}
}

@article{jm_zuo_1990,
  author  = {Zuo, Jian-Min and Rez, Peter},
  title   = {Observation of the structural phase transition in SrFio3 by diffuse electron scattering},
  journal = {Proceedings of the Electron Microscopy Society of America Annual Meeting},
  volume  = {48},
  pages   = {420--421},
  year    = {1990},
  month   = {08},
  doi     = {10.1017/S0424820100175235}
}

@article{PhysRevB.100.144305,
  author    = {Niermann, Tore},
  title     = {Scattering of fast electrons by lattice vibrations},
  journal   = {Phys. Rev. B},
  volume    = {100},
  number    = {14},
  pages     = {144305},
  year      = {2019},
  month     = {10},
  doi       = {10.1103/PhysRevB.100.144305}
}

@article{Nellist_1995,
  author  = {Nellist, P. D. and McCallum, B. C. and Rodenburg, J. M.},
  title   = {Resolution beyond the 'information limit' in transmission electron microscopy},
  journal = {Nature},
  volume  = {374},
  number  = {6523},
  pages   = {630--632},
  year    = {1995},
  month   = {04},
  doi     = {10.1038/374630a0}
}

@article{rodenburg_noise,
  author  = {Godard, Pierre and Allain, Marc and Chamard, Virginie and Rodenburg, John},
  title   = {Noise models for low counting rate coherent diffraction imaging},
  journal = {Optics express},
  volume  = {20},
  number  = {23},
  pages   = {25914--25934},
  year    = {2012},
  month   = {11},
  doi     = {10.1364/OE.20.025914}
}

@article{Hage_EELS,
author = {F. S. Hage  and G. Radtke  and D. M. Kepaptsoglou  and M. Lazzeri  and Q. M. Ramasse },
title = {Single-atom vibrational spectroscopy in the scanning transmission electron microscope},
journal = {Science},
volume = {367},
number = {6482},
pages = {1124-1127},
year = {2020},
doi = {10.1126/science.aba1136},
}

@article{Yan_EELS,
	Author = {Yan, Xingxu and Liu, Chengyan and Gadre, Chaitanya A. and Gu, Lei and Aoki, Toshihiro and Lovejoy, Tracy C. and Dellby, Niklas and Krivanek, Ondrej L. and Schlom, Darrell G. and Wu, Ruqian and Pan, Xiaoqing},
	Da = {2021/01/01},
	Doi = {10.1038/s41586-020-03049-y},
	Journal = {Nature},
	Number = {7840},
	Pages = {65--69},
	Title = {Single-defect phonons imaged by electron microscopy},
	Ty = {JOUR},
	Volume = {589},
	Year = {2021}
    }

@article{cos_metric_1,
	author = {Otsuka, Yanosuke},
	journal = {Bulletin of the Biogeographical Society of Japan},
	pages = {165--170},
	title = {The faunal character of the Japanese Pleistocene marine Mollusca, as evidence of the climate having become colder during the Pleistocene in Japan},
	year = {1936},
    Ty = {JOUR},
     volume={6},
    number = {16}
	}

@article{Cowley1957_Multislice,
    author = {J. M. Cowley AND A. F. Moodie},
    title = {The Scattering of Electrons by Atoms and Crystals. I. A New Theoretical Approach},
    journal = {Acta Cryst. A},
    year = 1957,
    volume = 10,
    pages = {609-619}
}

@article{Fischer88_Uequiv,
    author = {R. X. Fischer and E. Tillmanns},
    year = 1988,
    journal={Acta Cryst. C},
    volume =44,
    pages={775--776},
    title = {The equivalent isotropic displacement factor},
}

@article{gao_review,
	Author = {Mao, Ruilin and He, Peiyi and Liu, Fachen and Shi, Ruochen and Du, Jinlong and Gao, Peng},
	Doi = {10.1021/acsnano.4c17750},
	Journal = {ACS nano},
	Month = {05},
	Title = {Electron Microscopy for Nanophononics: A Review},
	Ty = {JOUR},
	Volume = {19},
	Year = {2025}}

@book{Kirkland_book,
	Author = {Kirkland, Earl},
	Date = {2010/06/25},
	Doi = {10.1007/978-1-4419-6533-2{\_}12},
	Isbn = {978-1-4419-6532-5},
	Month = {06},
	Title = {Advanced Computing in Electron Microscopy},
	Year = {2010}}
\end{document}


\maketitle  

\section*{Experimental Conditions}

The 4D-STEM dataset of hexagonal boron nitride bicrystal was acquired in a Nion HERMES microscope using a single 256x256 pixel wide chip of Dectris ELA direct electron detector. The acquisition time, beam current, accelerating voltage, beam convergence semi-angle and the real-space scan step were set to $2$ ms, $19$ pA, $60$ kV, $40$ mrad and $30.4$ pm, respectively. Diffraction patterns were recorded with a reciprocal sampling of $0.61$ mrad per detector pixel. In total, 7 identical 4D-STEM datasets were acquired one after another at the same region of interest and a non rigid registration \cite{nonrigid} was applied to eliminate the sample drift. One of these 7 datasets was used to perform the ptychographic reconstruction presented in this paper. In order to exclude inelastic scattering we further employed energy-filtering. In order to correct the geometric distortion created by the energy filter in the recorded diffraction patterns we acquired two tilt series with a small convergence semi-angle using the ELA detector (placed after the energy filter) and a detector placed before the filter. Using two tilt datasets we fitted a distortion vector field and applied an inverse transformation to the recorded diffraction patterns.

\section*{Theory}
Strictly speaking, an iterative ptychographic algorithm fits a forward model that, for a given scanning position (two coordinates $\rho_{p,x}$ and $\rho_{p,y}$) maps an illumination wavefront to a measured diffraction pattern given in terms of  spatial frequencies $k_{x}$ and $k_{y}$. This model includes a transmission function of the investigated sample and can be formulated in various complexity levels. Recovering the initially unknown transmission function is the main goal of any ptychographic algorithm, as its amplitude characterizes absorption while its phase is directly proportional to the specimen's electrostatic potential. During the reconstruction one can additionally refine the probe \cite{ePIE, paper_rop, ADORYM}, scan positions  \cite{paper_rop, ADORYM, maiden_positions} or a mis-tilt angle between the optical axis of the microscope and the zone axis of the studied crystal \cite{tilt_correction}.

When an electron beam  passes through a sufficiently thin sample, it experiences only one scattering event. The three dimensional structure of an object can be simplified to two lateral dimensions by integrating over the beam propagation direction. For a beam position $\rho_p$, the exit wave $\psi^{(exit)}(\rho_p, \rho)$ becomes a real space product of a two-dimensional wave function of the incident beam $\psi^{(in)}(\rho-\rho_p)$ with a two-dimensional complex transmission function of a specimen $O(\rho)$. The corresponding diffraction pattern (far-field intensity) can be calculated as an absolute squared modulus of the Fourier transformed exit wave:
\begin{align}
I(\rho_p, k)=\left |\mathcal F \left \{ \psi^{(exit)}(\rho_p, \rho) \right \}\right|^2=\left |\mathcal F \left \{ \psi^{(in)}(\rho-\rho_p)\cdot O(\rho)  \right \}\right|^2\label{eq: intensity}.
\end{align}

Increasing the beam wavelength and specimen thickness makes the effect of multiple scattering more pronounced. \cite{muller_vibrations} showed that at some point the thin object approximation starts to fail. In this case the most efficient strategy is to "divide and conquer". Instead of using one two dimensional transmission function, one can split the propagation direction into multiple intervals and define a set of 2D transmission functions responsible for each particular sufficiently thin region. We can write

\begin{align}
    \psi^{(exit)}_j(\rho_p, \rho)=\psi^{(in)}_j(\rho-\rho_p)\cdot O_j(\rho),
\end{align}
 where $j$ indicates a particular interval, i.e. slice. The propagation between the neighboring slices $j$ and $j+1$ over the interval $d$ is computed using a convolution with a Fresnel propagator:
\begin{align}
    &\psi^{(in)}_{j+1}(\rho)=\mathcal{F}^{-1}\left\{ \mathcal{F}\left\{\psi^{(exit)}_j(\rho) \right \} \cdot \mathcal{P}_{Fr}\left(k\right)\right \}\\
   & \mathcal{P}_{Fr}\left(k\right)=exp\left[-i\pi\lambda d |k|^2
\right] \label{eq: propagator},
\end{align}
\noindent
where $\lambda$ is a wavelength of the electron beam and the equation \ref{eq: propagator} defines the Fresnel propagator in reciprocal space. Typically one chooses the distance between the slices of approximately 1 nm  \cite{muller_vibrations, marcel_3D_recon}. The $\psi^{(in)}_{j=0}(\rho-\rho_p)$ is the incident illumination wavefront and for N slices the exit-wave $\psi^{(exit)}_{j=N}(\rho)$ is used to calculate a diffraction pattern as described in the equation \ref{eq: intensity}. Often the beam propagation direction slightly deviates from the zone axis of a crystal. Then, the Fresnel propagator can be modified \cite{Kirkland_book, tilt_correction} to compensate misalignment angles up to a few degrees. For two misalignment angles $\alpha_x$ and $\alpha_y$ along $x$ and $y$ axes, respectively, the tilted Fresnel propagator is defined as follows:
\begin{equation}
    \mathcal{P}_{Fr}\left(k, \alpha_x, \alpha_y\right)=exp\left[-i\pi \left(\lambda d |k|^2 + 2 k_x \alpha_x + 2 k_y \alpha_y\right) \right] \label{eq: tilt_prop}.
\end{equation}

In a real experimental situation, it is not always appropriate to neglect the partial spatial coherence of the electron source and vibrations of the atoms. To account for partial spatial coherence, Thibault and Menzel \cite{Thibault_mixed_state} proposed to replace the pure probe state $\psi^{(in)}_{j=0}(\rho)$  with a statistical mixture of multiple probe states $\psi^{(in)}_{j=0, m}(\rho)$, where the first index $j=0$ remained from the multi-slice formalism and the second index $m$ accounts for multiple modes. The total predicted diffraction pattern is calculated as an incoherent sum of the intensities corresponding to the individual probe modes. Let $I^{(1)}(\psi^{in}_{j=0}(\rho-\rho_p),O(\rho))
$ denote the sequence of operations required to obtain a diffraction pattern from a single probe mode. In mixed-probe formalism, the intensity is modeled as
\vspace{-0.25cm}
\begin{align}
I_{total}=\frac{1}{N_{probe\ modes}}\sum_{m=0}^{N_{{probe\ modes}}} I^{(1)}(\psi^{in}_{j=0, m}(\rho-\rho_p),O(\rho)).
\end{align}

The mixed-object formalism \cite{Thibault_mixed_state} is a natural extension of the mixed-probe formalism that accounts for a non-stationary object's transmission function. Due to the lattice vibrations and the corresponding displacements of the atoms, two electrons hitting the specimen at exactly the same spatial position but at two different points in time interact with slightly different electrostatic potentials. To account for this effect, i.e. thermal diffuse scattering (TDS) \cite{loane_TDS_1991}, one can use multiple transmission functions and model a diffraction pattern as an incoherent average of the intensities corresponding to the individual pure transmission functions.
\vspace{-0.25cm}
\begin{align}
I_{total}=\frac{1}{N_{object\ modes}\cdot N_{probe\ modes}}\sum_{n=0}^{N_{{object\ modes}}} \sum_{m=0}^{N_{{probe\ modes}}} I^{(1)}(\psi^{in}_{j=0, m}(\rho-\rho_p),O_{n}(\rho)). \label{eq: master_mixed_equation}
\end{align}

Thus, in the most complex scenario one has to deal with a three dimensional illumination wavefront (2 lateral dimensions plus one dimension for multiple modes), a four dimensional object (2 lateral dimensions and two dimensions one each for multiple slices and multiple modes) and perform $N_{{object\ modes}} \times N_{{probe\ modes}}$ forward multi-slice propagations to model one diffraction pattern. 

\section*{Reconstruction Algorithm}

We employ an in-house written code based on a python library CuPy \cite{cupy_learningsys2017}. All reconstructions were done on a single NVIDIA H100 GPU. Our ptychographic reconstruction is organized as a gradient-based minimization of a metric \cite{paper_rop, ADORYM, lsq_ml}, i.e. a loss function, describing the discrepancy between the measured intensities $I^{m}$ and the ones predicted by the forward model, as defined in the master equation \ref{eq: master_mixed_equation} and further denoted as $I$. This study employs a Gaussian-likelihood metric \cite{rodenburg_noise}:
\begin{align}
    \mathcal{L}_{Gauss}=\sum \left(\sqrt{I^{m}}-\sqrt{I}\right)^2,\label{eq: gauss_error}
\end{align}
where the sum goes over all beam positions and all pixels.

The loss derivative and corresponding update-vectors, are computed via Wirtinger calculus \cite{Wirtinger} a limited-memory BFGS (l-BFGS)  algorithm \cite{Wirtinger, l-bfgs}. For simulated silicon data no additional regularization was employed and an initial guess for the object was created using uniform prior. The reconstruction show in Fig. 2e of the main text went in two stages: 1) A prior reconstruction with single object mode and 9 probe modes initialized using Hermite polynomials \cite{muller_vibrations}; 2) Final reconstruction using a probe fitted during the first stage with 20 object modes. This scheme allows to apply a constraint on the recovered correlations by absorbing as much collective motion of the beam or sample during the first stage and recover the non-uniform vibrations during the second one.

The experimental hBN dataset was padded with zeros in reciprocal space up to a shape of $732\times732$ pixels resulting in a pixel size of ptychographic reconstruction of $10.89$ pm. We further applied a $2/3$ frequency cutoff to all intermediate waves (i.e. between each slice) to prevent aliasing artifacts \cite{Kirkland_book}. The reconstruction went in four stages:
1) initial reconstruction with single probe mode, single object mode and 30 object slices. This allowed us to estimate the aberrations of the beam. Further, cross-correlating the slices with their next neighbors allowed us to determine a minuscule difference between the zone axis of the crystal and the beam propagation direction. We used this value and adopted a tilted propagator \cite{tilt_correction} as defined in \ref{eq: tilt_prop} without an additional refinement of the angle for all subsequent steps. 
2) reconstruction with 10 probe modes and a single object mode without any regularization using the angle fitted during the stage 1; 3) 10 object modes were generated by adding uniform prior to the object recovered at the stage 2; 4) After the object at stage 3 was free from noise we added at the beginning, we applied regularization to the object, in particular, an $l_1$-regularization \cite{paper_rop} of the absorption potential enforcing the amplitude of the retrieved transmission function to be closer to 1 and a missing-wedge regularization \cite{tilt_correction, muller_vibrations} damping the high $k_z$ Fourier-components at small $k_x$ and $k_y$.

\section*{Undersampling of the Beam Propagation Direction}

\begin{figure}[H]
\centering
\includegraphics[width=0.9\textwidth]{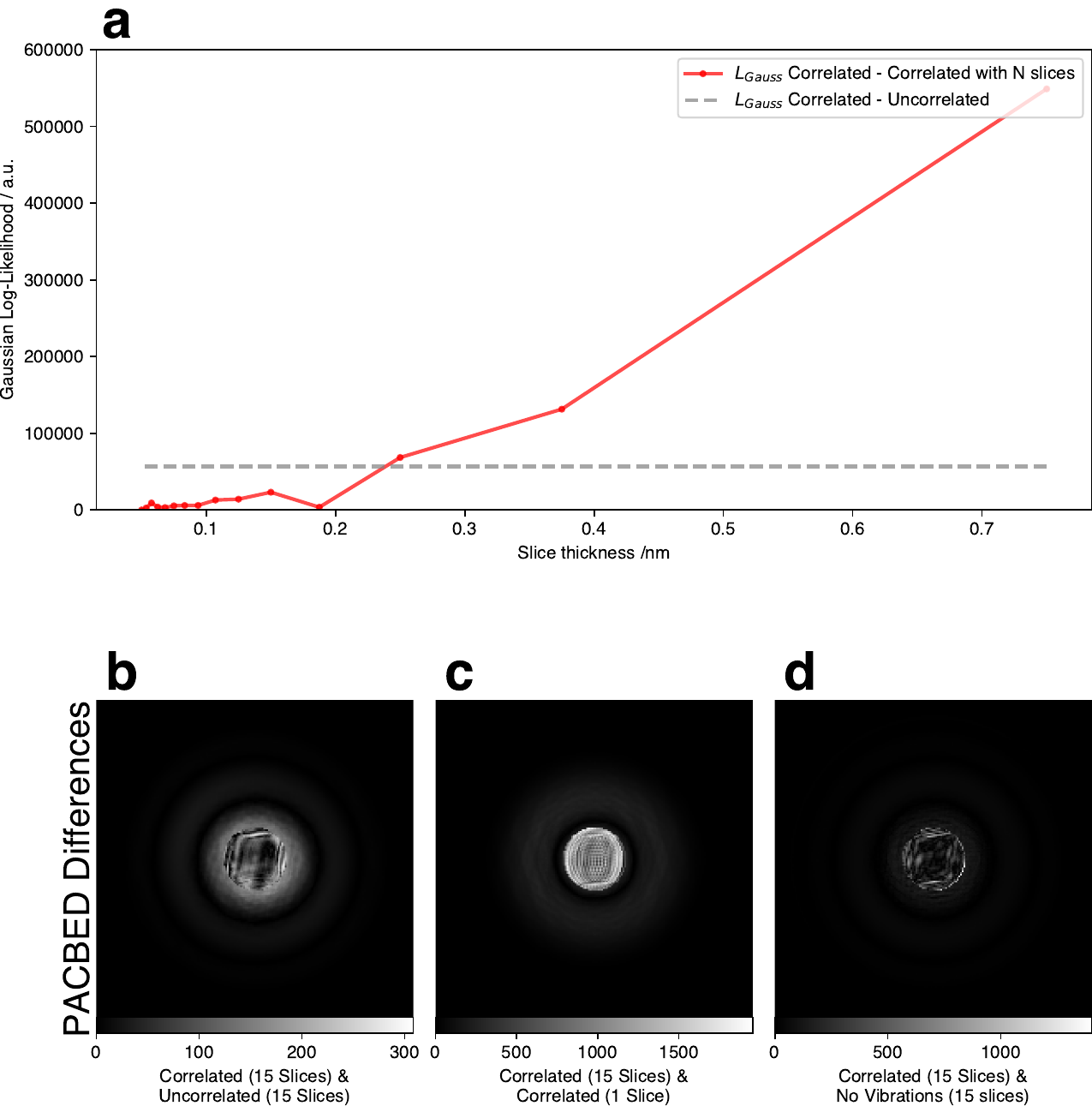}
\caption{\textbf{Effect of depth undersampling, i.e. of not sufficient slices along the beam propagation direction.} Panel \textbf{a} shows the Gaussian log-likelihood between a fully sampled correlated 4D-STEM dataset (0.5 \AAs thick slices) of a symmetric $\Sigma9$ grain boundary in silicon and multiple undersampled correlated datasets with increasing slice thickness (red), as well as between the correlated and uncorrelated datasets (gray dashed). Panels \textbf{b-–d} show PACBED difference maps comparing: \textbf{b} correlated vs uncorrelated, \textbf{c} correlated vs correlated-undersampled (15 slices), and \textbf{d} correlated vs absent vibrations data. Artifacts from undersampling exceed those caused by interatomic vibrations. Thus, to recover vibrations one should first account for effect of multiple scattering,  even with a 0.7 nm thick sample.}
\label{fig: plot_slices}
\end{figure}

We would like to highlight an importance of accounting for multiple scattering when performing mixed-object ptychography. Reconstructions presented in the main text were done using slice thickness of 2 \AA, but the corresponding simulations at 0.5 \AA. The reason for such fine sampling in the reconstruction lies in the fact, that the effect of under-sampling along the beam propagation direction may be much more noticeable than the features produced by a particular type of atomic motion. We performed additional 4D-STEM simulations using same 30 MD snapshots of the silicon grain boundary but with an increased slice thickness.  In \figref{fig: plot_slices} we show a gaussian-log likelihood (in arbitrary units) between one correlated dataset with slice thickness of 0.5 \AAs and another correlated dataset with a larger slice thickness. The gray dashed line shows Gaussian log-likelihood between correlated and uncorrelated datasets, both simulated using slice thickness of 0.5 \AA. From the plot clearly follows that the difference between various kinds of interatomic vibrations is far weaker than the difference caused by under-sampling along the beam propagation direction. Further in \figref{fig: plot_slices}[b -- d] we show differences between position-averaged diffraction patterns (PACBED) of correlated and uncorrelated datasets, correlated and undersampled-correlated datasets, and correlated and perfectly coherent datasets, respectively. 

\printbibliography

\section*{Extended data figures}

\begin{figure}[H]
\centering
\includegraphics[width=\textwidth]{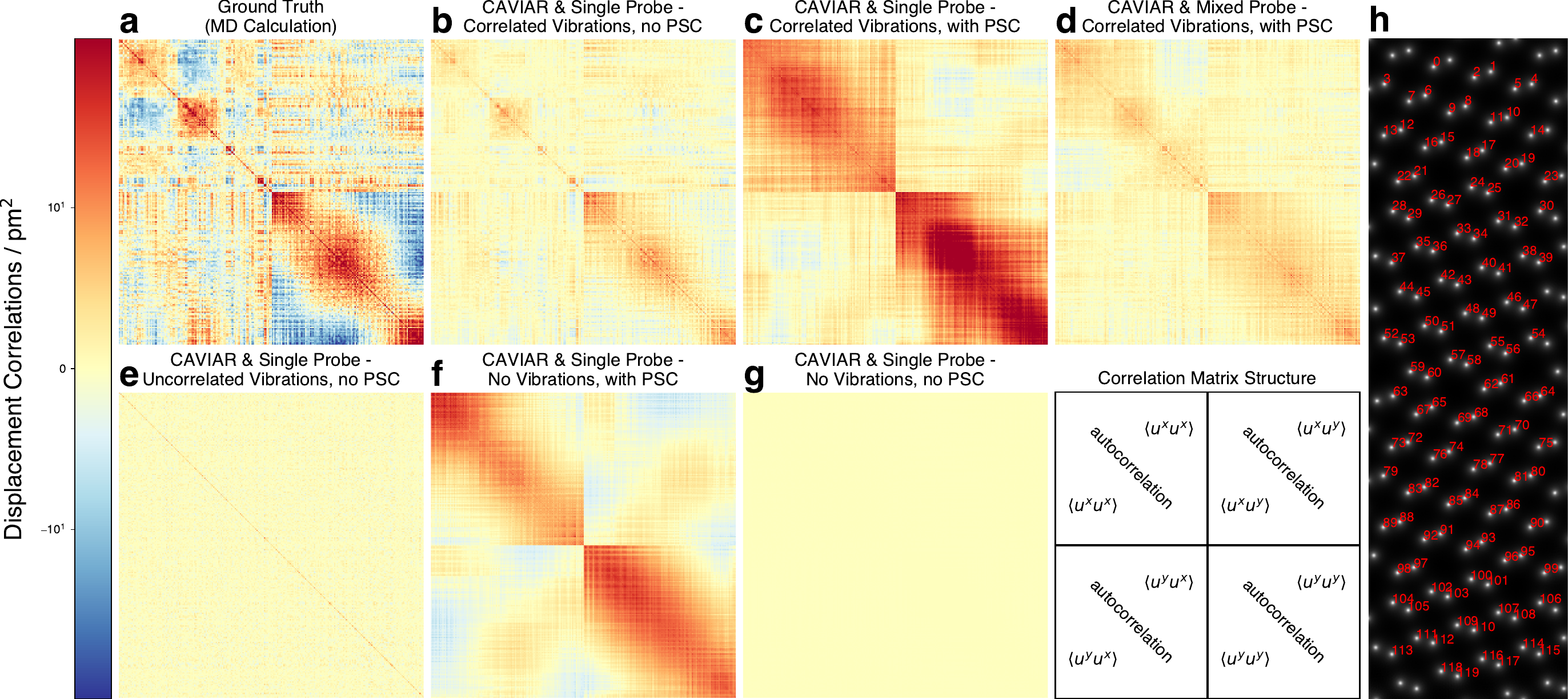}
\caption{
\textbf{Full Green's tensors obtained via CAVIAR from various 4D-STEM datasets of a symmetric $\Sigma$9 grain boundary in silicon.} \textbf{a,} The ground truth correlation matrix derived from molecular dynamics simulations with clear off-diagonal structure indicating spatially correlated atomic displacements. \textbf{b,} The result of CAVIAR reconstruction from the dataset with correlated vibrations and no partial spatial coherence (PSC), preserving the main features of the ground truth but with slightly lower magnitude. \textbf{c,} CAVIAR reconstruction using a single probe mode from correlated dataset with additional partial spatial coherence, producing a structure similar to Panel \textbf{a} but offset by a uniform background. \textbf{d,} CAVIAR reconstruction from the same data as in \textbf{c}, but using 9 probe modes. Mixed probe efficiently mitigates the mentioned offset in correlations and produces correlations very similar to the ones shown in \textbf{a} and \textbf{b}. \textbf{e,} The reconstruction from a dataset based on the Einstein model without additional PSC, off-diagonal elements of the Green's tensor are close to zero indicating an absence of correlations between different degrees of freedom. \textbf{f,} A reconstruction from a dataset without any atomic vibrations but with partial spatial coherence (the dataset was not presented in the main text of the paper). The observed correlations stem from the incoherent illumination rather than atomic motion illustrating a positive offset mentioned the main text of the paper. \textbf{g,} The result for a perfectly static sample without PSC, where the correlation matrix exhibits near-zero values throughout, reflecting the absence of motion. In panel \textbf{h} we show indexing of atoms on top of a ground-truth phase. The structure of the matrices is schematically shown in the lower right corner. All CAVIAR reconstructions \textbf{b -- g} except \textbf{d} used a single probe mode, in \textbf{d}, 9 incoherent probe modes were used.}\label{fig: 4slices_corrs_full}
\end{figure}

\begin{figure}[H]
\centering
\includegraphics[width=\textwidth]{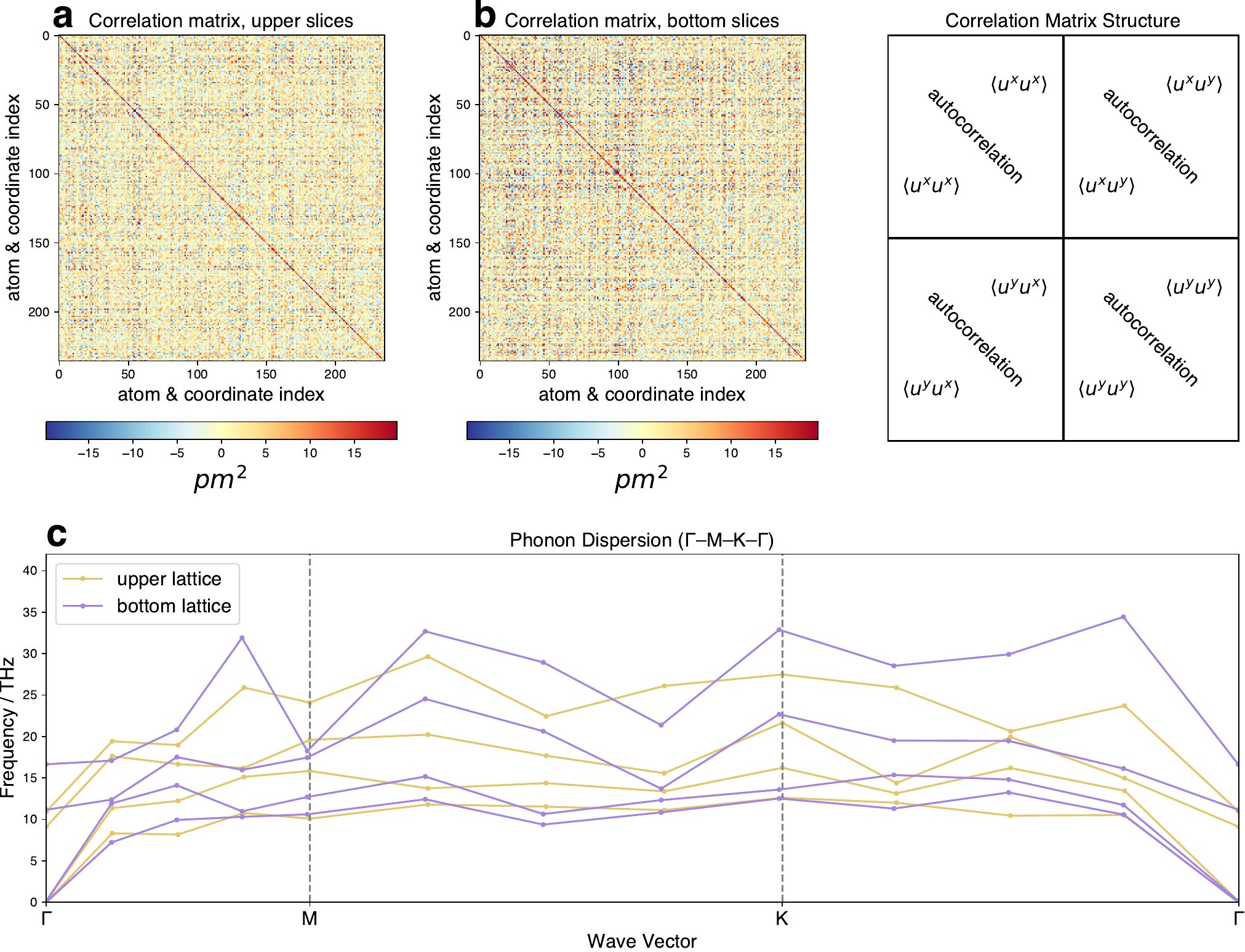}
\caption{
\textbf{Interatomic correlations (Green's tensors) recovered from experimental mixed-object ptychographic reconstruction of approximately $15$ nm thick bulk hBN crystal.} Images \textbf{a} and \textbf{b} show average correlation matrices extracted from  a $z$-range $z=1.9$ nm $\ldots 5.7$ (top lattice) and $z=8.6$ nm $\ldots 12.5$ nm (bottom lattice), respectively. The indices of the matrices run over all atoms in available in the field of view and two spatial coordinates ($x$, $y$), the structure of the matrices is schematically depicted on the right side. \textbf{c}, Phonon dispersion curves extracted from the reconstructed atomic displacement correlations using the lattice Green’s tensor method from an experimental bulk hBN dataset. Yellow curves show the dispersion from the upper hBN lattice layer, and purple curves the one from the lower layer. Both are computed along the $\Gamma$-M–K-$\Gamma$ direction of the Brillouin zone. The curves reveal acoustic and optical branches, although their absolute frequencies are reduced due to limited sampling of object states, limited spatial resolution and the lack of $z$-displacement sensitivity in the 2D projection. In the upper lattice the average frequencies along the $\Gamma-K-M-\Gamma$ path are $10.7$ THz, $14.1$ THz,  $17.7$ THz and $24.0$ THz for transverse acoustic, longitudinal acoustic, transverse optical, longitudinal optical branches in upper lattice. For bottom lattice we get  $10.8$ THz, $13.0$ THz, $18.2$ THz, and $27.0$ THz for the same four branches.
}\label{fig: G_tensor_hbn}
\end{figure}